\documentclass{eccomas2016}
\usepackage{amsmath}
\usepackage{amsfonts}
\usepackage{amssymb}

\title{ON THE VALIDITY OF TIDAL TURBINE ARRAY CONFIGURATIONS OBTAINED FROM STEADY-STATE ADJOINT OPTIMISATION}

\author{Christian T. Jacobs$^1$, Matthew D. Piggott$^1$, Stephan C. Kramer$^1$ and Simon W. Funke$^2$}

\heading{Christian T. Jacobs, Matthew D. Piggott, Stephan C. Kramer and Simon W. Funke}

\address{$^1$Department of Earth Science and Engineering, Imperial College London \\
  London SW7 2AZ, United Kingdom\\
  e-mail: \{c.jacobs10,m.d.piggott,s.kramer\}@imperial.ac.uk \and
  $^2$
  Biomedical Computing Department, Simula Research Laboratory \\
  P.O. Box 134, 1325 Lysaker, Norway\\
  e-mail: simon@simula.no}


\abstract{Extracting the optimal amount of power from an array of tidal turbines requires an intricate understanding of tidal dynamics and the effects of turbine placement on the local and regional scale flow. Numerical models have contributed significantly towards this understanding, and more recently, adjoint-based modelling has been employed to optimise the positioning of the turbines in an array in an automated way and improve on simple man-made configurations (e.g. structured grids of turbines) \cite{Funke_etal_2014}.

Adjoint-based optimisation of high-resolution and ideally 3D transient models is generally a very computationally expensive problem. Multiple approaches are therefore used in practice to obtain feasible runtimes: using high viscosity values to obtain a steady-state solution, or a sequence of steady-state solutions for ``time-varying'' setups; limiting the number of adjoint computations; or reformulating the problem to allow for coarser mesh resolution to make it feasible for resources assessment (e.g. \cite{Funke_etal_Submitted}, \cite{Culley_etal_2015}). However, such compromises may affect the reliability of the modelled turbines, their wakes and interactions, and thus bring into question the validity of the computed optimal turbine positions.

This work considers a suite of idealised simulations of flow past tidal turbine arrays in a two-dimensional channel. It compares four regular array configurations, detailed by Divett et al. \cite{Divett_etal_2013}, with the configuration found through adjoint optimisation in a steady-state, high-viscosity setup. The optimised configuration produces considerably more power than the other configurations (approximately 40\% more than the best man-made configuration). The same configurations are then used to produce a suite of transient simulations that do not use constant high-viscosity, and instead use large eddy simulation (LES) to parameterise the resulting turbulent structures. All simulations are performed using OpenTidalFarm \cite{Funke_etal_2014}.

It is shown that the `low background viscosity'/LES simulations produce less power than that predicted by the constant high-viscosity runs. Nevertheless, they still follow the same trends in the power curve throughout time, with optimised layouts continuing to perform significantly better than simplified configurations.}

\begin{document}

\section{INTRODUCTION}

Tidal stream turbines are a promising new source of renewable energy. In order for the deployment of turbines to be effective and economically viable, their placement in a turbine array/farm must be carefully considered. Knowledge of the tidal flow conditions, the effect the turbines will have on the flow and any natural habitats, and how any resulting changes will affect turbine power generation downstream (e.g. through blockage effects), all need to be investigated. Experimental studies are often limited to small-scale domains with idealised flow conditions, which may not scale well to realistic marine environments. As more computational resources become available to solve bigger problems, numerical modelling and adjoint-based optimisation of tidal turbine positions is becoming an increasingly popular alternative.

Current work on adjoint-based optimisation has shown very promising results with regards to improving the efficiency of turbine farms over simple man-made configurations, such as a structured grid of turbines or staggering them in some regular way. For example, Funke et al. \cite{Funke_etal_2014} optimised an array in the Inner Sound of the Pentland Firth, comprising 256 turbines, which resulted in a 33\% increase in power extraction. Similar array optimisation studies by Culley et al. \cite{Culley_etal_2014} and Barnett et al. \cite{Barnett_etal_Submitted} in idealised scenarios and once again in the Pentland Firth yielded increased power generation of up to 25\%. However, despite the power that adjoint-based optimisation offers, it is extremely computationally-demanding. Such optimisation requires both the forward run and the adjoint run (i.e. the simulation run backwards in time) to be performed at every iteration of an optimisation procedure, based on numerical algorithms such as L-BFGS-B (the limited-memory Broyden-Fletcher-Goldfarb-Shanno (BFGS) algorithm with bound constraints) \cite{Byrd_etal_1995, Zhu_etal_1997}. High-resolution transient simulations of flow past marine power turbines, particularly those that feature turbulent wakes and thus require even smaller time-steps to capture the physics, are often prohibitively expensive.

As a result, current work on the adjoint optimisation of tidal turbine placement in the literature has been mostly limited to steady-state simulations in which very high, uniform and constant, non-physical values of the background viscosity are required to ensure that a steady-state solution exists (e.g. 0.5 m$^2$s$^{-1}$ \cite{Funke_etal_Submitted}, 3 m$^2$s$^{-1}$ \cite{Funke_2012, Barnett_etal_Submitted}, 30 m$^2$s$^{-1}$ \cite{Culley_etal_2014, Culley_etal_2015, Funke_etal_2014}, 90 m$^2$s$^{-1}$ \cite{Culley_etal_2016}). Such simulations only require one solve of the governing forward equations (with the time derivatives removed) and one solve of the adjoint equations\footnote{The simulation can also be divided up into $n$ steady-states, in which case 2$n$ solves are required per optimisation iteration.}, per optimisation iteration, making the problem tractable with available resources. However, the wake downstream of each turbine is likely to be diffused out in such simulations, and no turbulence is modelled. The validity of the results from the optimisation process is therefore questionable. The size of the wake and the presence of any turbulent eddies can potentially affect the power generated by the turbines further downstream, and any blockage effects from neighbouring turbines may not be realistic. When the optimised turbine positions from the steady-state computations are applied to a realistic, transient flow comprising a much lower background viscosity and higher Reynolds number, the amount of power generated by the flow may be substantially different and the benefits of the optimisation process may not be as pronounced compared to the steady-state, high-viscosity case.

The work presented in this paper first considers four simple man-made turbine array configurations by Divett et al. \cite{Divett_etal_2013}. Steady-state simulations are run with each of these configurations, and the configuration that provides the highest amount of generated power is then optimised using adjoint-based optimisation. The resulting optimised configuration, and also the four man-made configurations, are then each considered in a transient setup in which a more realistic background viscosity is used and Large Eddy Simulation (LES) is applied to parameterise the turbulence. Section \ref{sect:model} presents the shallow water equations that are solved by the numerical model, along with details of the Smagorinsky LES turbulence parameterisation that is enabled in the transient flow simulations. Setup details are provided in Section \ref{sect:setup}, followed by an analysis of the steady-state and transient simulation results in Section \ref{sect:results}. The paper closes with some concluding remarks in Section \ref{sect:conclusions}.

\section{NUMERICAL MODEL}\label{sect:model}
This work uses the OpenTidalFarm numerical modelling package \cite{Funke_etal_2014} for solving the shallow water equations. It uses FEniCS \cite{Logg_etal_2012} to discretise the equations with the finite element method, and the Dolfin-Adjoint framework \cite{Farrell_etal_2013, FunkeFarrell_Submitted} to automatically annotate the forward model system and compute the adjoint model based on that.

\subsection{Momentum equation}
The equation governing conservation of momentum is given by
\begin{eqnarray}\label{eq:momentum}
  \frac{\partial \mathbf{u}}{\partial t} + \mathbf{u}\cdot\nabla\mathbf{u} = -g\nabla \eta + \nabla\cdot\mathbb{T} - \frac{c_b + c_t}{H}\|\mathbf{u}\|\mathbf{u},
\end{eqnarray}
where $\mathbf{u}$ is the depth-averaged velocity, $g$ is the acceleration due to gravity, $\eta$ is the free-surface displacement, $h$ is the free-surface height at rest, $H$ is the total free-surface height (i.e. $h + \eta$), $c_b$ is the (dimensionless) bottom drag coefficient, and $c_t$ is the (dimensionless) tidal turbine drag coefficient. The Euclidean norm $\|\mathbf{u}\| = \sqrt{\mathbf{u}\cdot\mathbf{u}}$ is used here. The stress tensor $\mathbb{T}$ is given by
\begin{eqnarray}\label{eq:stress_tensor}
  \mathbb{T} = \nu\left(\nabla\mathbf{u} + \nabla\mathbf{u}^{\mathrm{T}}\right),
\end{eqnarray}
where $\nu$ is the kinematic viscosity.

Each of the $N$ turbines is represented by a two-dimensional Gaussian profile, such that for a turbine $i$ of radius $r_i$ centred at ($x_i$, $y_i$),
\begin{eqnarray}
  c_t = \sum_{i=1}^N K_i\phi_{x_i,r_i}(x)\phi_{y_i,r_i}(y),
\end{eqnarray}
where $K_i$ is a dimensionless friction coefficient and
\begin{eqnarray}
  \phi_{p,r}(x) = \left\{
	\begin{array}{ll}
		\exp(1-1/(1-|\frac{x-p}{r}|^2))  & \mbox{if } |\frac{x-p}{r}| < 1 \\
		0 & \mbox{otherwise. }
	\end{array}
\right.
\end{eqnarray}

\subsection{Continuity equation}
The continuity equation is given by
\begin{eqnarray}\label{eq:continuity}
  \frac{\partial \eta}{\partial t} + \nabla\cdot\left(H\mathbf{u}\right) = 0.
\end{eqnarray}

\subsection{Turbulence parameterisation}
Large eddy simulation (LES) is used to parameterise the turbulence generated in the transient flow simulations with a low background viscosity. The crux of LES is to resolve the large-scale eddies by using a fine enough mesh spacing, whilst modelling any smaller, sub-grid scale eddies with an eddy viscosity term which acts as a diffusive agent; it essentially describes the energy transfer from the larger (resolved) eddies down to the smaller eddies. The particular LES model employed in this work is the standard Smagorinsky model \cite{Smagorinsky_1963, Deardorff_1970}. The eddy viscosity term is given by
\begin{eqnarray}\label{eq:eddy_viscosity}
 \nu^{\prime} = \left(c_s\Delta\right)^2|\mathbb{S}|,
\end{eqnarray}
where $c_s$ is the Smagorinsky coefficient, set to 0.2 in this work (within the range of typical $c_s$ values \cite{Deardorff_1971}). The filter width $\Delta$ is an estimate of the local element size which is defined here as the square root of the area of each element. The strain rate tensor $\mathbb{S}$ is defined as
\begin{eqnarray}
 \mathbb{S} = \frac{1}{2}\left(\nabla\mathbf{u} + \nabla\mathbf{u}^{\mathrm{T}}\right),
\end{eqnarray}
The modulus of $\mathbb{S}$ is given by
\begin{eqnarray}
|\mathbb{S}| = \sqrt{2\sum_{i}\sum_{j}{\mathbb{S}_{ij}\mathbb{S}_{ij}}}.
\end{eqnarray}
where $\mathbb{S}_{ij}$ is the ($i$,$j$)-th component of $\mathbb{S}$. The eddy viscosity $\nu^{\prime}$ is added to the background viscosity $\nu$ in (\ref{eq:stress_tensor}) and is updated at the beginning of each time-step.

Turbulence is a truly three-dimensional phenomenon. Furthermore, LES requires a substantial amount of resolution to sufficiently resolve the large-scale eddies. The validity of applying an LES model in this work to a two-dimensional domain is therefore questionable. Nevertheless, 2D turbulence modelling in the context of tidal turbine simulation has shown good agreement with experimental data \cite{Nash_etal_2015}, and similar LES schemes such as MILES have generated promising results in resolving turbine wakes \cite{Divett_etal_2014, Divett_etal_2016}.

\section{SIMULATION SETUP}\label{sect:setup}
All the simulations presented here are two-dimensional and were performed in a 3 $\times$ 1 km$^2$ rectangular domain (in the $x$-$y$ plane), following a similar setup to the work by Divett et al. \cite{Divett_etal_2013}. A total of 15 turbines were deployed in the four man-made configurations as described in \cite{Divett_etal_2013}, and a suite of steady-state shallow water simulations was first performed for each of the configurations. The physical parameters are provided in Table \ref{table:parameters}.

 \begin{table}[h]
   \begin{center}
     \begin{tabular}{*{2}{c}}
     \hline
     Parameter & Value \\
     \hline
     Free-surface height at rest & $h$ = 50 m \\
     Kinematic viscosity (steady-state) & $\eta$ = 1 m$^2$s$^{-1}$ \\
     Kinematic viscosity (transient) & $\eta$ = $10^{-6}$ m$^2$s$^{-1}$ \\
     Smagorinsky coefficient & $c_s$ = 0.2 \\
     Gravitational acceleration & $g$ = 9.81 ms$^{-2}$ \\
     Water density & $\rho$ = 1,000 kgm$^{-3}$ \\
     Bottom drag coefficient & $c_b$ = 0.0025 \\
     Number of turbines & $N$ = 15 \\
     Turbine radii & $r_i$ = 10 m $\forall$ $i$ = 1, $\ldots$, $N$\\
     Turbine friction coefficient & $K_i$ = 12 $\forall$ $i$ = 1, $\ldots$, $N$\\
     \hline
     \end{tabular}
   \end{center}
 \caption{The common parameter values used in all simulations. The Smagorinsky coefficient $c_s$ is only required in the transient simulations where the Smagorinsky LES model is used. Note also the difference in the background kinematic viscosity between the steady-state and transient simulations.}
 \label{table:parameters}
 \end{table}

\subsection{Mesh}
The computational mesh that discretised the 3 $\times$ 1 km$^2$ domain, comprising triangular elements, was generated using Gmsh \cite{GeuzaineRemacle_2009}. The inner region defined by 500 $\leq x \leq$ 2,500 m and 125 $\leq y \leq$ 875 m, where all the turbines are situated, contained a structured grid of solution nodes with a characteristic element length of $\Delta x$ = 2 m. The outer section of the domain's mesh, which is outside the region of interest and therefore does not require high numerical resolution, was unstructured with $\Delta x$ = 100 m.

\subsection{Initial and boundary conditions}
The initial conditions $\mathbf{u}(x, t=0) = [0, 0]^\mathrm{T}$ ms$^{-1}$ and $\eta(x, t=0) = 0$ m were applied at $t = 0$ s. Throughout the simulations a Dirichlet velocity boundary condition of 2 ms$^{-1}$ was strongly imposed to allow inflow from the left. For the steady-state simulations, outflow was modelled by a zero-value Dirichlet free-surface condition strongly imposed along the right-hand boundary. For the transient simulations, a Flather \cite{Flather_1976} condition was used to minimise spurious reflections at the outlet. No-normal flow conditions were weakly imposed along the lateral walls.

\subsection{Spatial and temporal discretisation}
The Galerkin finite element method was used to discretise the model equations in space. Continuous piecewise quadratic Lagrange basis functions were used to represent the velocity solution field, whereas continuous piecewise linear Lagrange basis functions were used for the free-surface $\eta$ (thereby forming the P2-P1, or Taylor-Hood, element pair \cite{LarsonBengzon_2013}) as well as all other fields (including $\nu$ and $\nu^\prime$).

Unlike the steady-state simulations, the transient simulations required temporal discretisation due to the presence of time derivatives. The implicit backward Euler method was used for this purpose so that larger time-steps could be taken. All transient simulations were run until $T$ = 2,000 s with a time-step of $\Delta t$ = 4 s, which gave ample opportunity for the turbulent flow to become fully developed and for any spurious reflections from the outflow to dissipate. Newton iteration dealt with the non-linearity introduced through the advection term in (\ref{eq:momentum}); for each time-step, $k$ iterations were performed such that $\|\mathbf{u}^{k+1} - \mathbf{u}^{k}\| \leq 10^{-9}$, where $k$ is typically between 3 and 10. Once both (\ref{eq:momentum}) and (\ref{eq:continuity}) are discretised and assembled to form a fully block-coupled system, the system was solved directly with LU decomposition. 

\subsection{Man-made turbine configurations}

\subsubsection{Centred}
\begin{figure}[ht]
  \begin{center}
    \includegraphics[width=10cm]{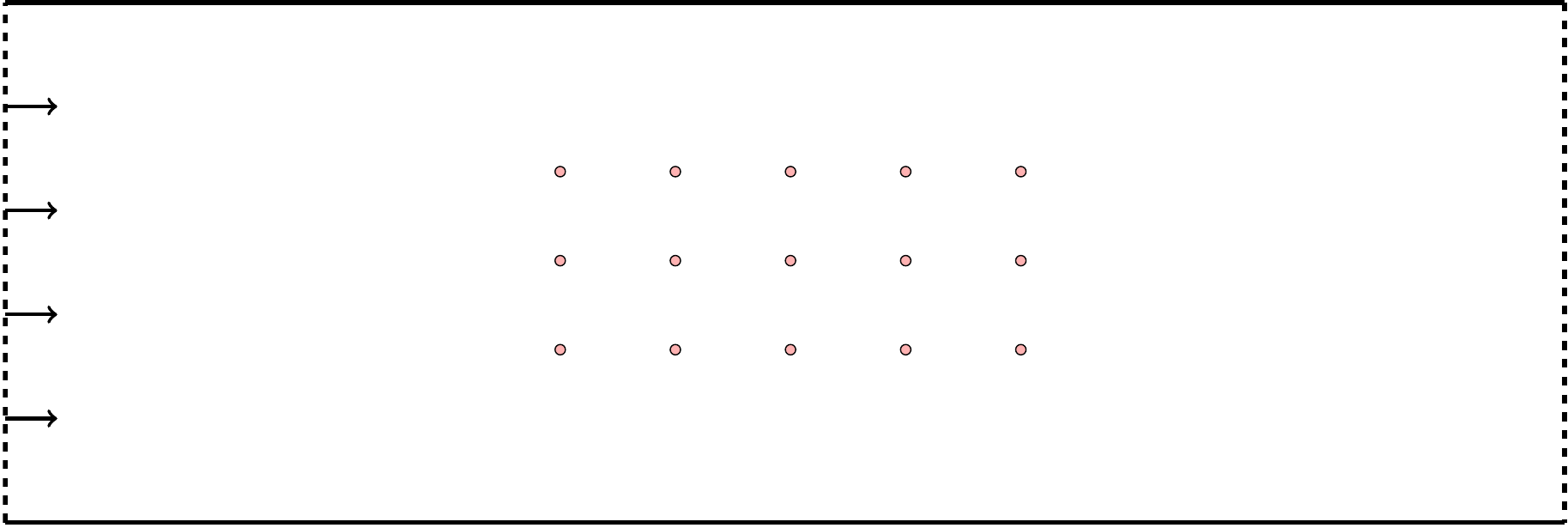}
    \caption{The centred turbine configuration. The whole 3 $\times$ 1 km$^2$ channel is shown.}
    \label{fig:centred}
  \end{center}
\end{figure}

As per the work in \cite{Divett_etal_2013}, the centred configuration illustrated in Figure \ref{fig:centred} considered a regular 5 $\times$ 3 grid of turbines positioned in the centre of the channel. The turbines in each row are 7.5$d$ away from each other, and each row is 10$d$ apart, where $d$=2$r$ is the turbine diameter.

\subsubsection{Offset}
\begin{figure}[ht]
  \begin{center}
    \includegraphics[width=10cm]{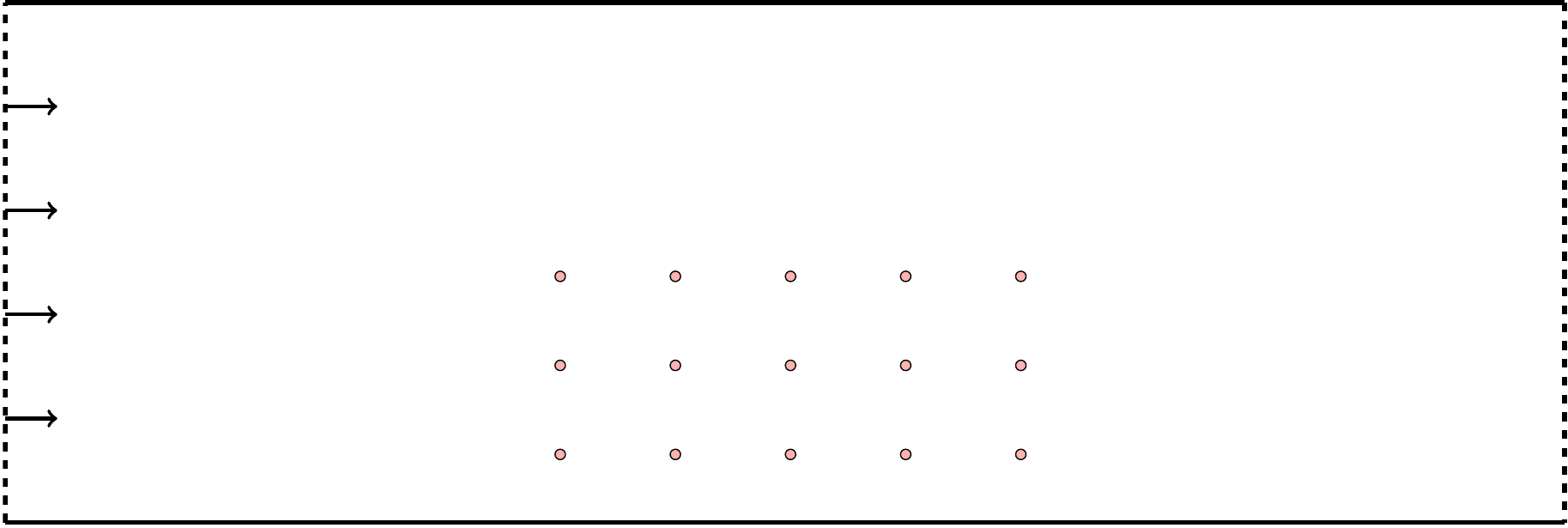}
    \caption{The offset turbine configuration. The whole 3 $\times$ 1 km$^2$ channel is shown.}
    \label{fig:offset}
  \end{center}
\end{figure}

The offset case (see Figure \ref{fig:offset}) considers the same regular grid of turbines as the centred case, but instead of being placed in the centre of the domain, the grid is positioned closer to one of the side walls. Such a configuration may be required to, for example, allow cargo to be transported via ship down one side of the channel \cite{Divett_etal_2013}.

\subsubsection{Staggered}
\begin{figure}[ht]
  \begin{center}
    \includegraphics[width=10cm]{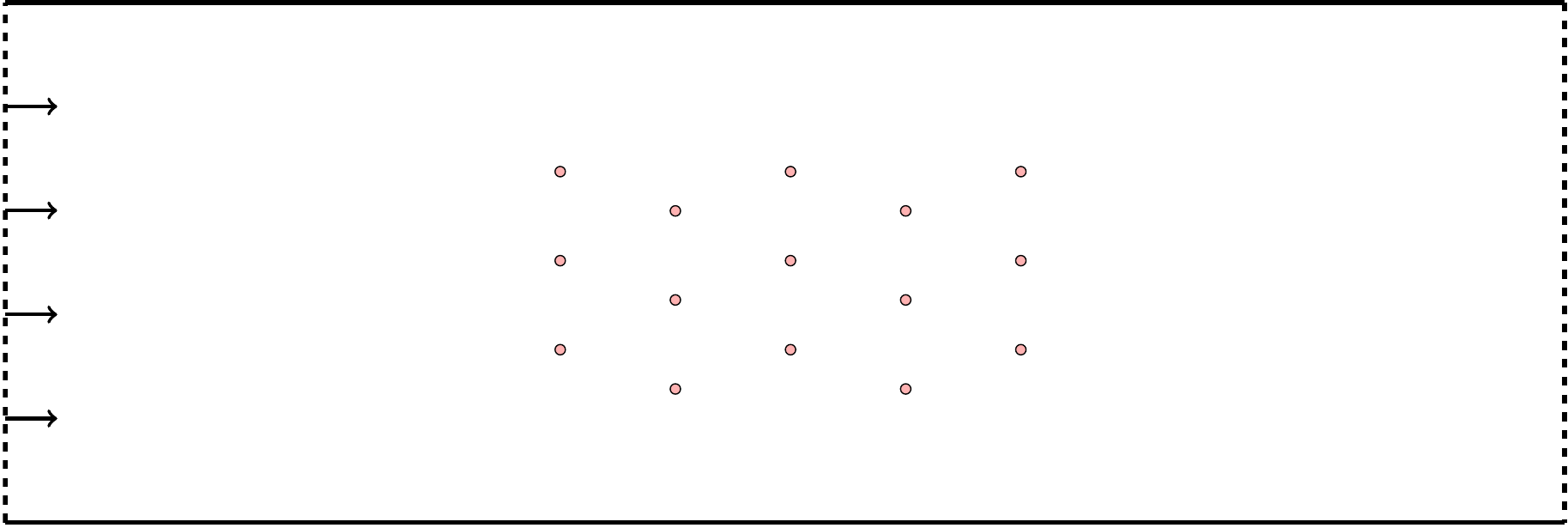}
    \caption{The staggered turbine configuration. The whole 3 $\times$ 1 km$^2$ channel is shown.}
    \label{fig:staggered}
  \end{center}
\end{figure}

This is similar to the centred configuration, but each row is staggered such that the turbines in one row are aligned between the turbines of the adjacent row. This is illustrated in Figure \ref{fig:staggered}.

\subsubsection{Larger spacing}
\begin{figure}[ht]
  \begin{center}
    \includegraphics[width=10cm]{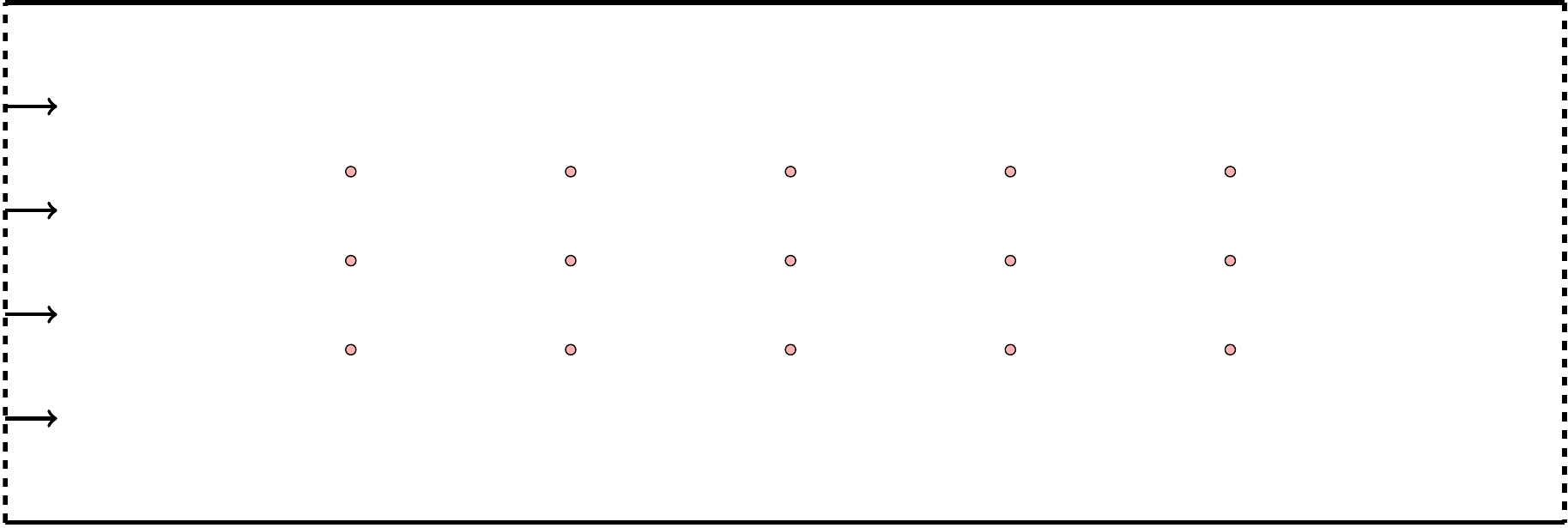}
    \caption{The larger spacing turbine configuration. The whole 3 $\times$ 1 km$^2$ channel is shown.}
    \label{fig:larger_spacing}
  \end{center}
\end{figure}

In the larger spacing case (see Figure \ref{fig:larger_spacing}), the spacing between each turbine in a given row remains the same as the centred case, but the spacing between each row is now 20$d$ instead of 10$d$.

\subsection{Adjoint optimisation}
Adjoint-based optimisation was employed to maximise the amount of power extracted by the turbines with respect to the individual turbine positions. Such optimisation of high-resolution transient simulations is computationally expensive. Therefore, in order for the adjoint problem to be feasibly solved with available resources, only the steady-state problem was optimised in this work.

The Dolfin-Adjoint library \cite{Farrell_etal_2013, FunkeFarrell_Submitted} was used to annotate the governing equations and automatically differentiate the forward model with respect to the optimisation variables, chosen to be the individual turbine positions here. The adjoint system was solved to obtain gradient information, which was then passed to the L-BFGS-B algorithm \cite{Byrd_etal_1995, Zhu_etal_1997} to optimise the functional describing the total power extracted by the turbine array, defined by

\begin{eqnarray}\label{eq:power}
 P(\mathbf{u}, \mathbf{m}) = \int_\Omega{\rho c_t \|\mathbf{u}\|^3} \ \mathrm{d}x,
\end{eqnarray}

where $\rho$ is the density of water (1,000 kgm$^{-3}$), $\mathbf{m}$ is the vector of optimisation variables (in this case, the coordinates of each turbine), and $\Omega$ denotes the whole domain \cite{Funke_etal_2014}.

\begin{figure}[ht]
  \begin{center}
    \includegraphics[width=10cm]{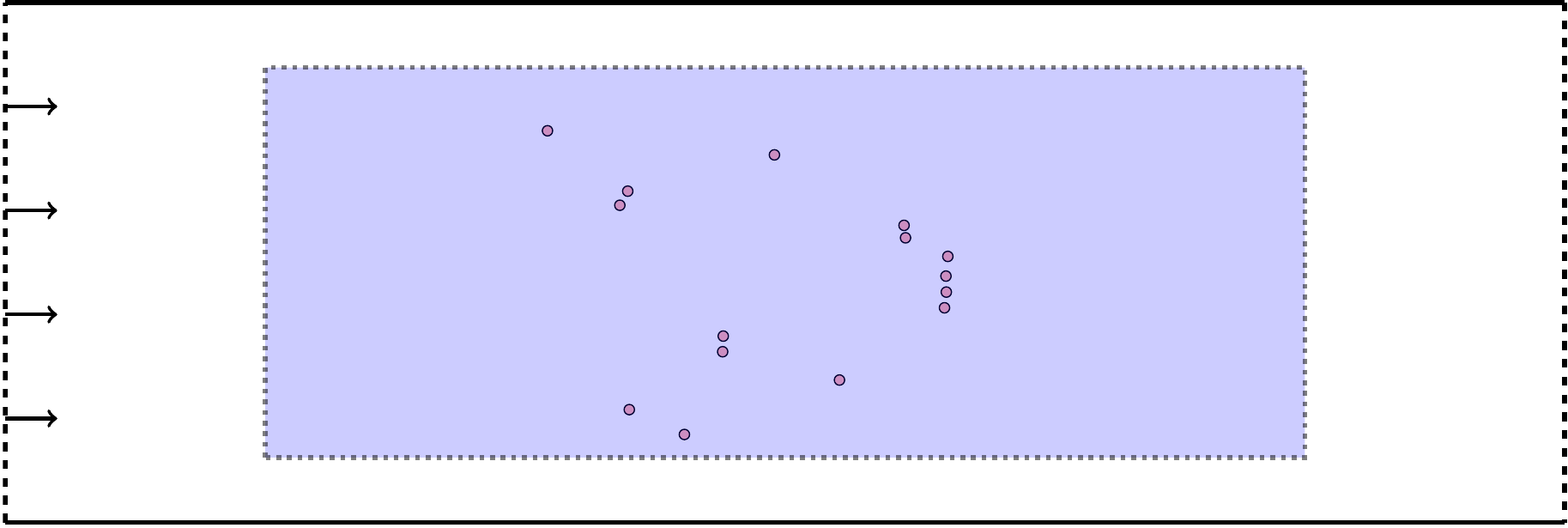}
    \caption{The optimised turbine configuration from the steady-state simulation after 100 optimisation iterations. The whole 3 $\times$ 1 km$^2$ channel is shown. The optimisation process was constrained such that turbines were not allowed to be placed outside of the dotted region shaded in blue. For the steady channel flow considered here a symmetric configuration would be expected, but this is not the case since the optimisation process was terminated after 100 iterations and had not fully converged (see Figure \ref{fig:optimisation_iterations}). Asymmetry may also be expected if the optimisation gets stuck in a local minimum.}
    \label{fig:optimised}
  \end{center}
\end{figure}

The staggered array configuration was used as the initial condition for the optimisation process, since this was found to be the best man-made configuration (see Section \ref{sect:results}). The optimisation was constrained such that turbines must be at least 1.5$d$ apart, where $d$ is the turbine diameter, and can only be placed in the region defined by 500 $\leq x \leq$ 2,500 m and 125 $\leq y \leq$ 875 m (i.e. the area of highest mesh resolution). The optimisation was terminated after 100 iterations, resulting in the configuration shown in Figure \ref{fig:optimised}.

\section{RESULTS}\label{sect:results}

\subsection{Steady-state simulations}
\begin{figure}[ht]
  \begin{center}
    \includegraphics[width=16cm]{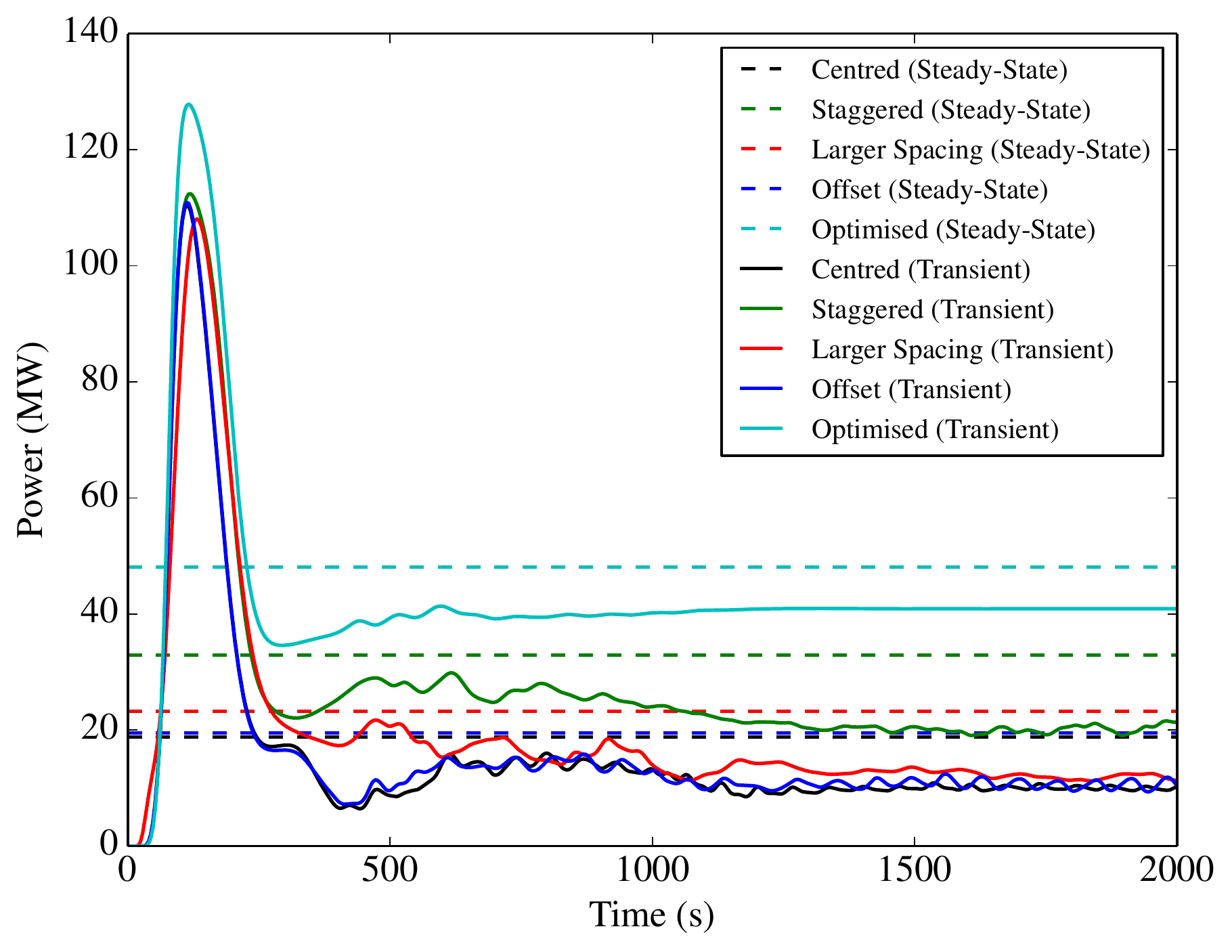}
    \caption{The total power generated against time, for each of the steady-state and transient simulations.}
    \label{fig:power}
  \end{center}
\end{figure}

\begin{figure}[ht]
  \begin{center}
    \includegraphics[width=16cm]{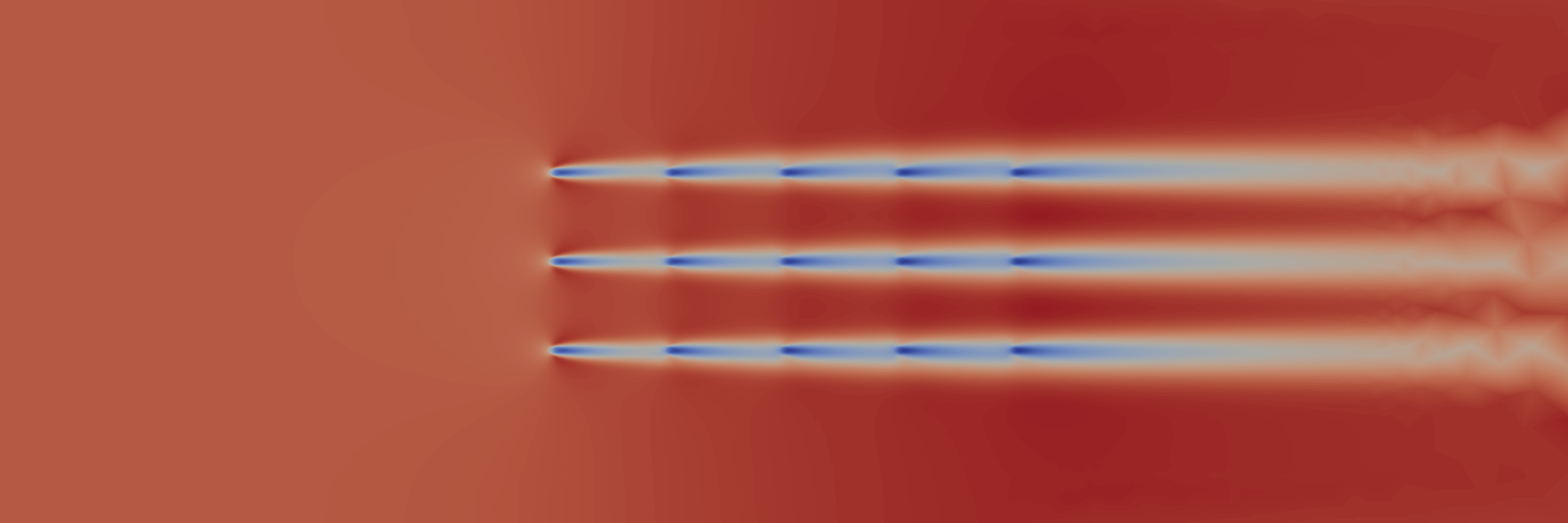}
    \includegraphics[width=8cm]{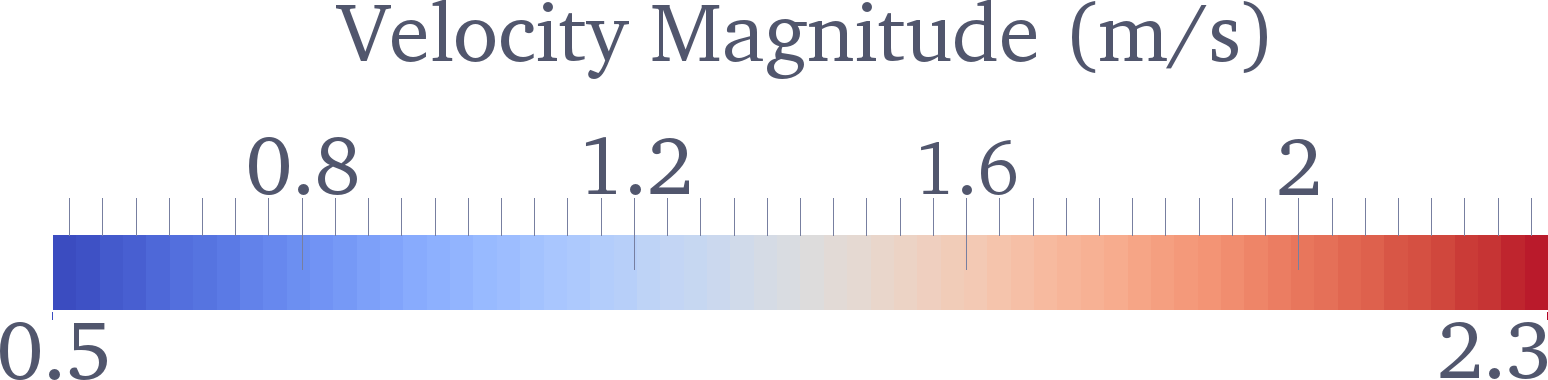}
    \caption{The flow speed around the centred turbine array in the steady-state simulation.}
    \label{fig:steady_centred}
  \end{center}
\end{figure}

\begin{figure}[ht]
  \begin{center}
    \includegraphics[width=16cm]{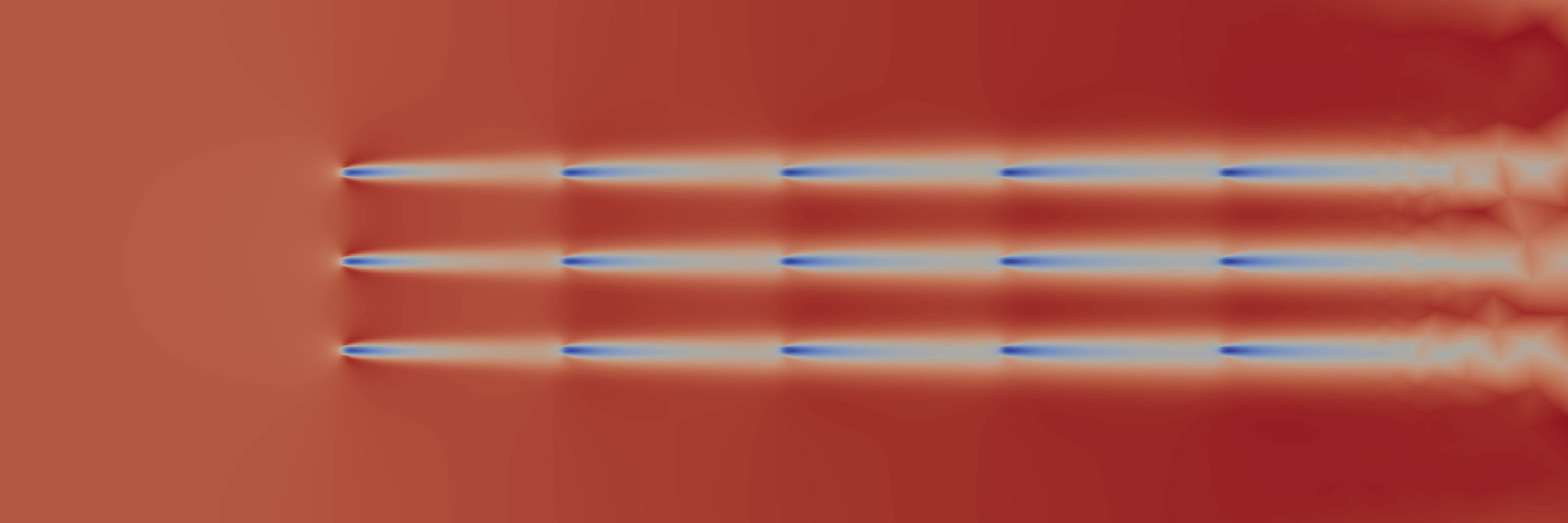}
    \includegraphics[width=8cm]{steady_legend.png}
    \caption{The flow speed around the array with larger turbine spacing in the steady-state simulation.}
    \label{fig:steady_larger_spacing}
  \end{center}
\end{figure}

\begin{figure}[ht]
  \begin{center}
    \includegraphics[width=16cm]{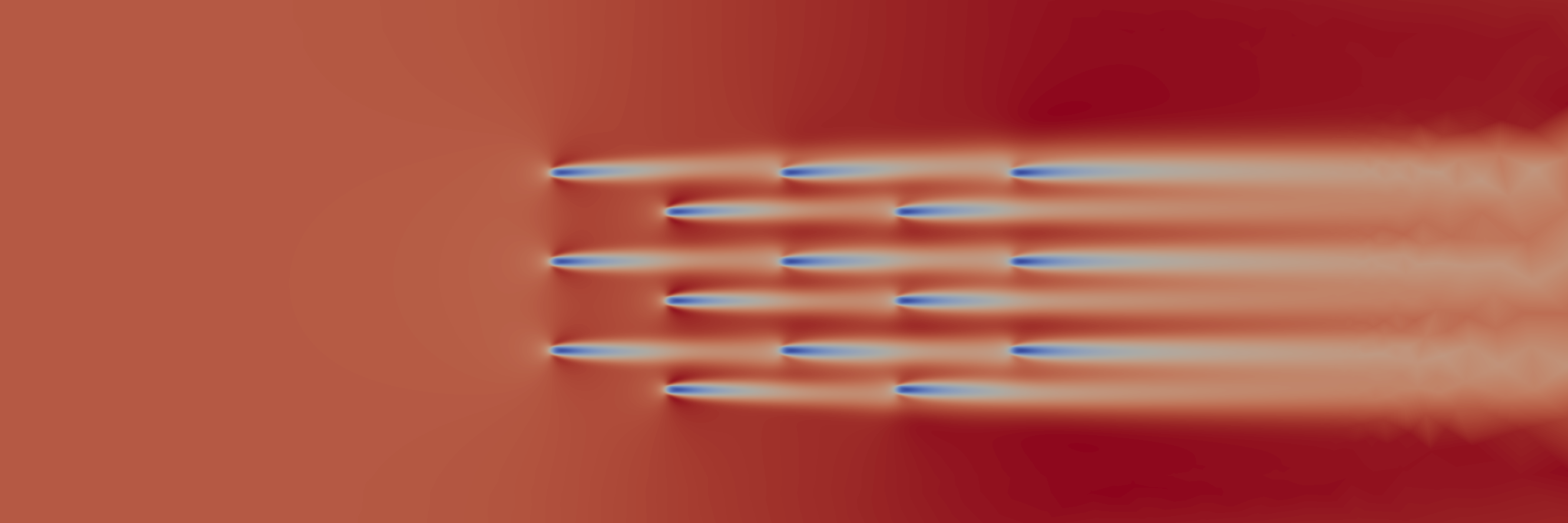}
    \includegraphics[width=8cm]{steady_legend.png}
    \caption{The flow speed around the staggered turbine array in the steady-state simulation.}
    \label{fig:steady_staggered}
  \end{center}
\end{figure}

The results from the steady-state simulations in Figure \ref{fig:power} show that the centred array was the poorest choice of configuration for this setup. The power extracted by such a setup was limited because, apart from the turbines in the first row, all the turbines downstream were situated in the wakes/separated flow regions of the ones upstream (as shown in Figure \ref{fig:steady_centred}). By definition there is a velocity deficit in these regions and thus, from equation (\ref{eq:power}), the power extracted was reduced. When a larger turbine spacing was used, the wake was able to recover before interacting with the downstream turbines (as shown in Figure \ref{fig:steady_larger_spacing}), and thus 23\% more power could be extracted from the flow. By offsetting the centred array configuration such that the turbines are closer to a side wall, the power extraction was increased, albeit fairly insignificantly, relative to the centred configuration. The staggered array configuration yielded the best power extraction potential, generating 75\% more power than the centred configuration. As Figure \ref{fig:steady_staggered} shows, the accelerated region of flow around the first row of turbines interacts with the second (staggered) row behind it, resulting in more energy being removed from the flow and thus more power extraction. Each of these findings agrees qualitatively with the transient simulations of \cite{Divett_etal_2013} which showed that, relative to the centred array, only a 4\% increase in power extraction could be obtained with the offset array, whereas a 31\% and 54\% increase could be obtained with the larger-spaced and staggered configurations, respectively.

\begin{figure}[ht]
  \begin{center}
    \includegraphics[width=16cm]{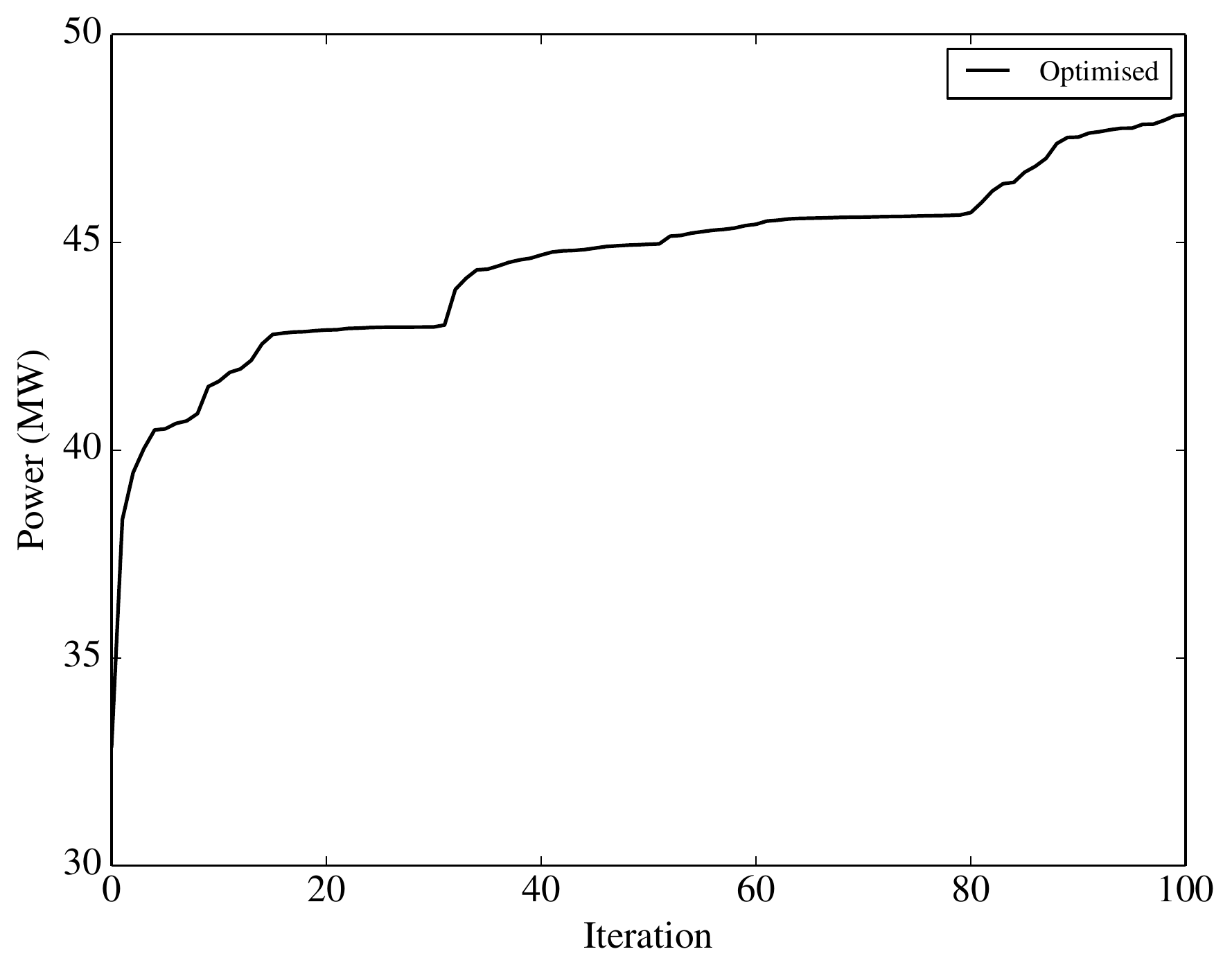}
    \caption{Total power generated at each adjoint optimisation iteration. Iteration 0 represents the power from the initial staggered configuration.}
    \label{fig:optimisation_iterations}
  \end{center}
\end{figure}

Since the staggered array yielded the most power, it was used as the initial condition and improved on throughout the adjoint optimisation process. This yielded a configuration that could extract 48.12 MW of power --- a 156\% increase relative to the centred array, as shown in Figure \ref{fig:optimisation_iterations}. This is greater than any difference in power extraction between the simple man-made configurations. Relative to the staggered array (i.e. the optimisation's initial condition and the best man-made configuration considered here), 46\% more power could be extracted.

\begin{figure}[ht]
  \begin{center}
    \includegraphics[width=16cm]{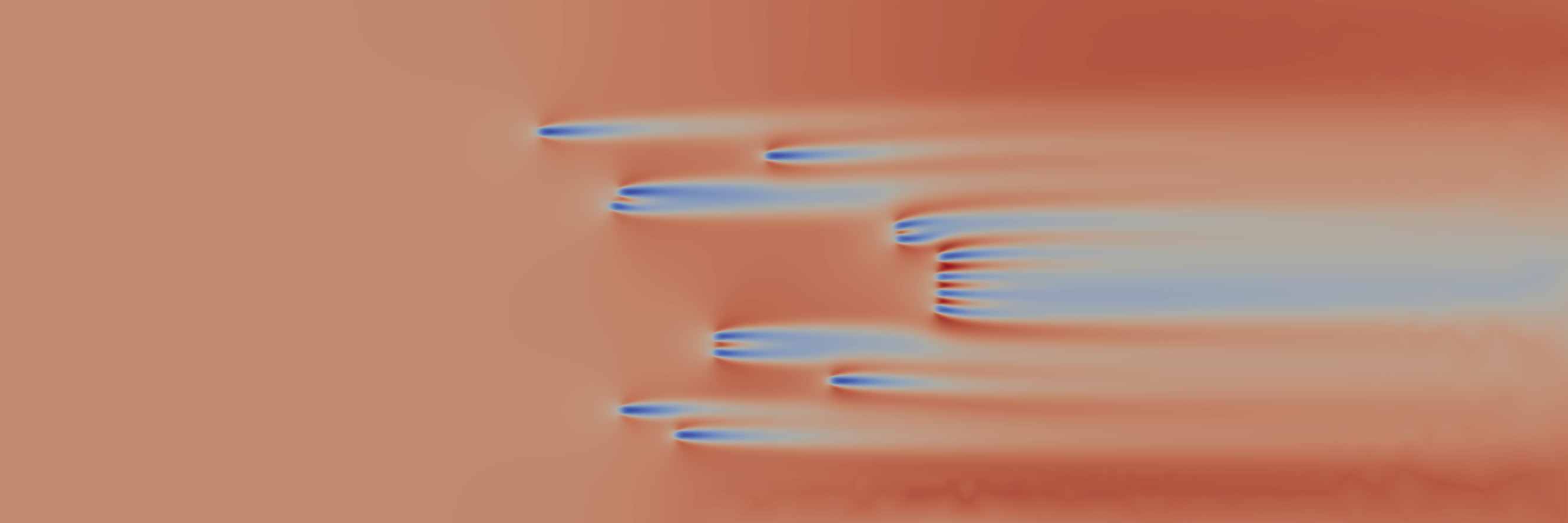}
    \includegraphics[width=8cm]{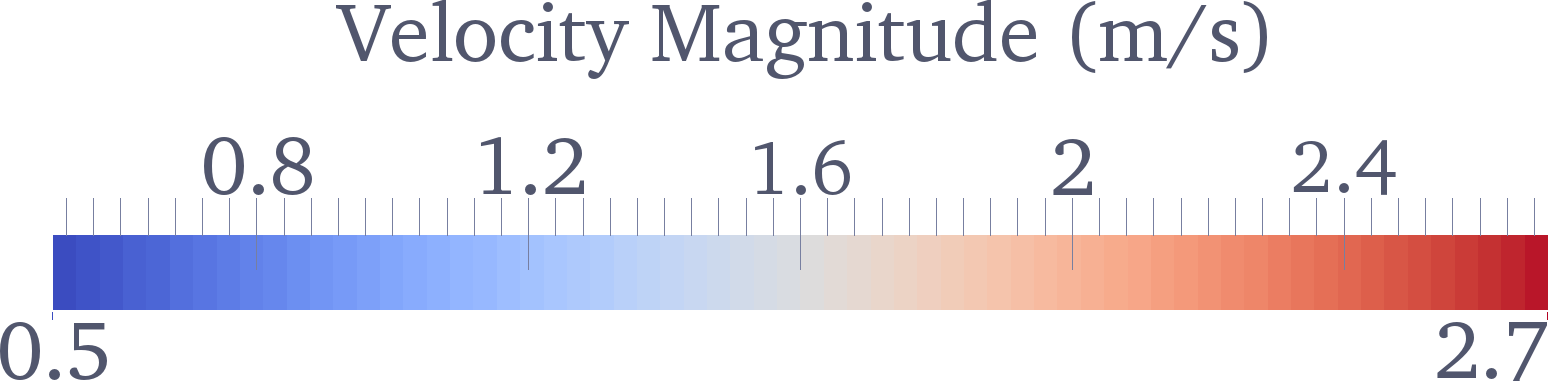}
    \caption{The flow speed around the optimised turbine array in the steady-state simulation. Note the increased upper limit of the colour bar.}
    \label{fig:steady_optimised}
  \end{center}
\end{figure}

Flow around a turbine results in acceleration around its edges, but also in a velocity deficit immediately behind the turbine (i.e. in its wake). It would be unwise to place a turbine in the wake of another since the power is related to the velocity. It therefore makes sense that the optimisation process re-positioned the turbines such that the effect of wakes from other turbines is minimised. Furthermore, regions of accelerated flow can be used to harness more power which is why some turbines are packed closer together in the cross-flow direction (e.g. the row of four turbines on the right-hand side of Figure \ref{fig:optimised}). This is highlighted in Figure \ref{fig:steady_optimised}.

\subsection{Transient simulations}
\begin{figure}[ht]
  \begin{center}
    \includegraphics[width=16cm]{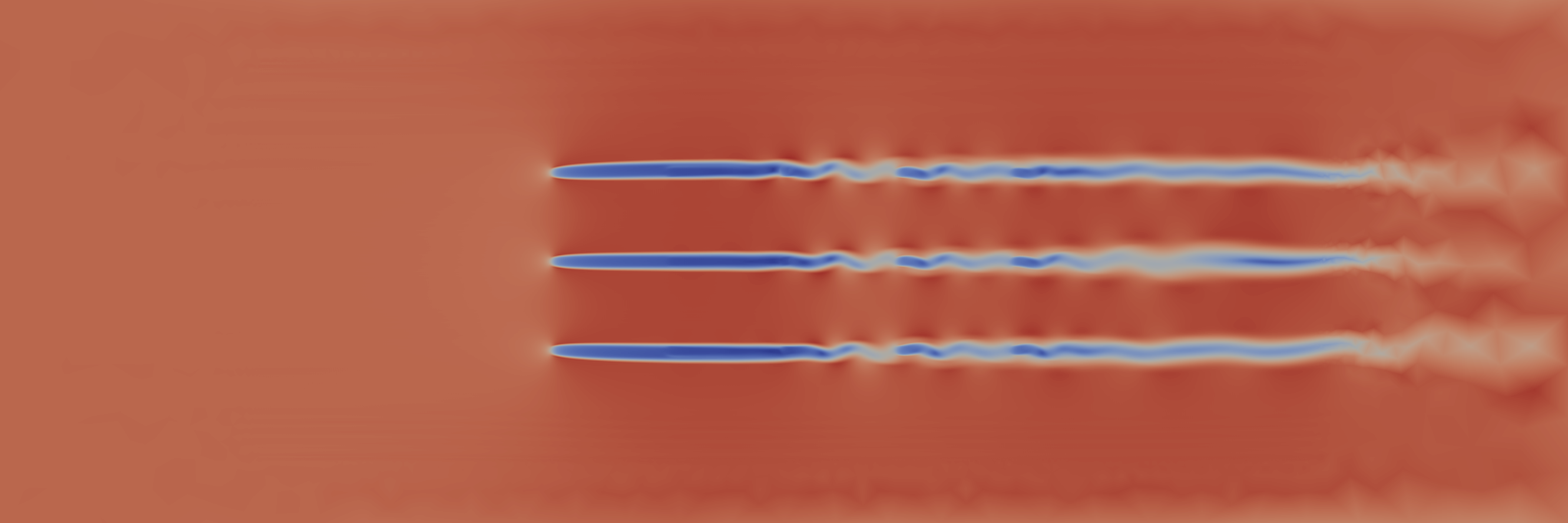}
    \includegraphics[width=8cm]{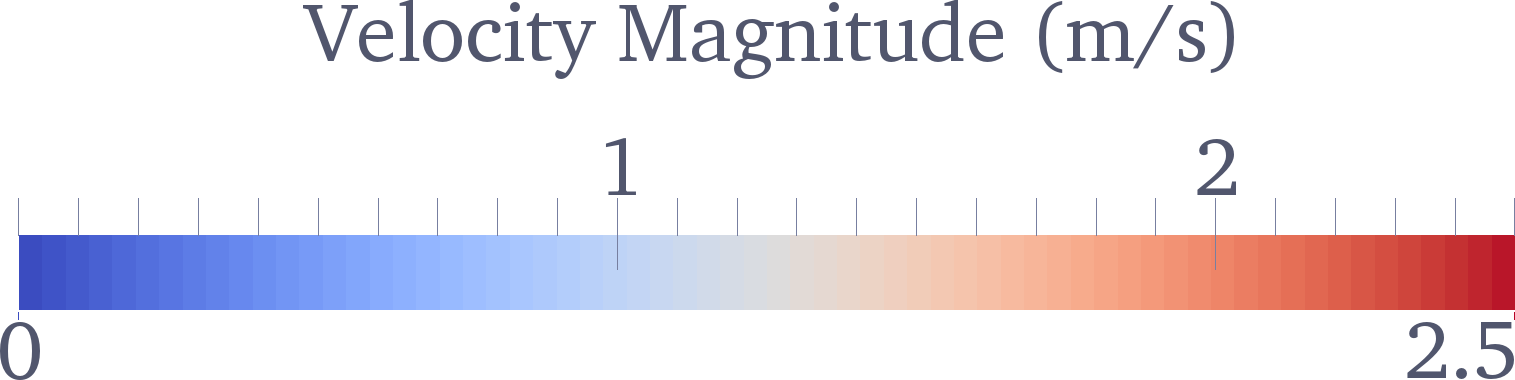}
    \caption{The flow speed around the centred turbine array in the transient simulation. Note the decreased lower limit of the colour bar.}
    \label{fig:transient_centred}
  \end{center}
\end{figure}

Unlike the steady-state simulations which featured highly dissipated wakes due to the presence of a uniformly high background viscosity, the transient simulations exhibited longer, sharper regions of separation with some downstream turbulent eddy shedding as Figure \ref{fig:transient_centred} demonstrates. From the power profiles in Figure \ref{fig:power}, the oscillations caused by turbulent eddies were relatively small. However, considerably less power was extracted compared to the steady-state setups, after the initial spin-up period; the initial peak was caused by the inflow condition of 2 ms$^{-1}$ superimposing itself with a region of 2 ms$^{-1}$ outflow generated by the Flather boundary condition, which eventually exited the domain. 

With respect to the differences in power between the configurations, qualitatively similar behaviour was observed. For example, in the centred configuration the turbine rows were still sufficiently close to be in the wakes of the upstream turbines. However, in the transient runs the velocity deficit was much stronger, dropping to a near-zero flow speed immediately behind the turbines, resulting in decreased power generation. In the simulation with larger turbine spacing, the improvement in power extraction relative to the centred array was not as significant compared to the steady-state setup, because the wakes were longer and sharper. The turbines should therefore be placed even further apart to allow the wake to sufficiently recover.

\begin{figure}[ht]
  \begin{center}
    \includegraphics[width=16cm]{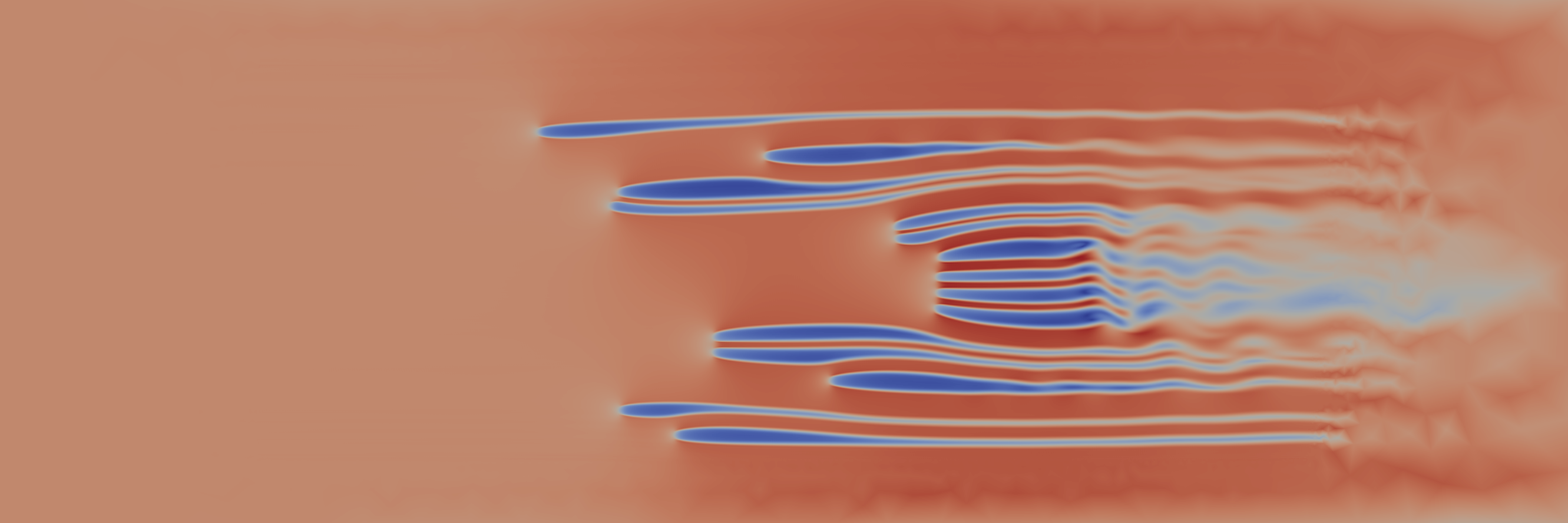}
    \includegraphics[width=8cm]{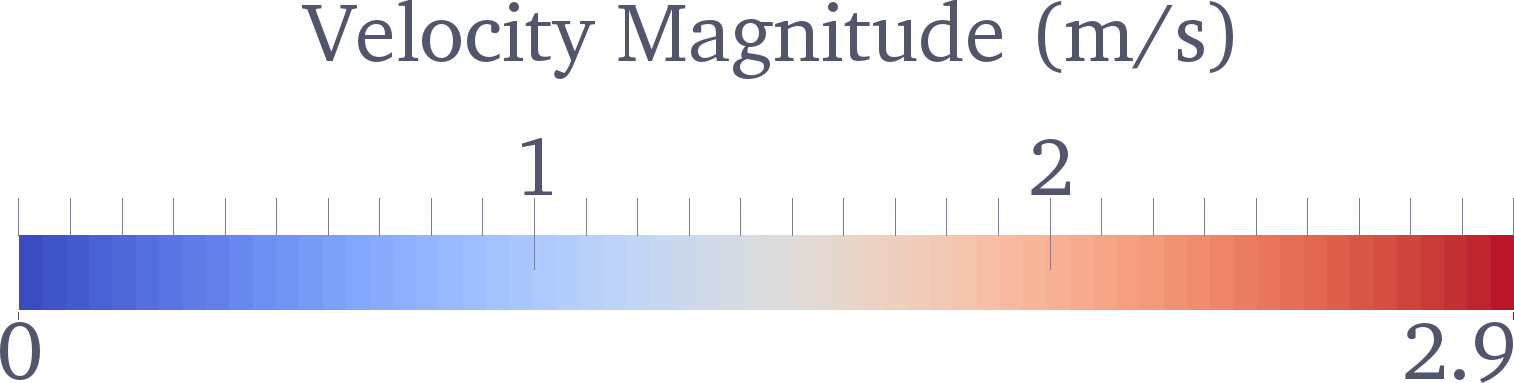}
    \caption{The flow speed around the optimised turbine array in the transient simulation. Note the increased upper limit of the colour bar as a result of increased flow speed around the turbines on the right-hand side that are situated very close together.}
    \label{fig:transient_optimised}
  \end{center}
\end{figure}

Once again, the optimised array configuration out-performed all four man-made configurations under consideration. Interestingly, many of the wakes were entrained into the regions of acceleration and flowed around the downstream turbines as Figure \ref{fig:transient_optimised} shows, instead of immersing them in the region of low flow speed and limiting power generation. Unlike the other configurations, turbulent eddies were only shed downstream of all the turbines, which explains why the power in Figure \ref{fig:power} remained constant once the flow became fully established.

\section{CONCLUSIONS}\label{sect:conclusions}

\begin{itemize}
  \item Adjoint optimisation can be used to greatly improve the power extraction potential of tidal turbine arrays/farms.
  \item The use of a lower background viscosity and turbulence modelling shows that the velocity deficit is much stronger in reality, resulting in decreased power generation relative to the high viscosity steady-state runs. Furthermore, turbulent eddies have the potential to affect the amount of power extracted, to a small degree in the configurations considered here.
  \item The power curves of the transient, turbulent runs are qualitatively similar to the steady-state runs, in that the best (staggered) man-made setup is still out-performed by the optimised configuration from adjoint modelling.
  \item Horizontal LES (HLES) \cite{Liek_2000} or 3D LES turbulence models, coupled with more realistic turbine parameterisations (e.g. rectangular blocks of drag as opposed to smooth Gaussian functions), should be adopted in future work.
\end{itemize}

\section{ACKNOWLEDGEMENTS}
The authors acknowledge the support and use of the Imperial College High Performance Computing Service and the UK National Supercomputing Service (ARCHER). CTJ was supported by ARCHER eCSE grant 03-7. SWF is supported by The Research Council of Norway through a Centres of Excellence grant to the Center for Biomedical Computing at Simula Research Laboratory, project number 179578. The authors would also like to acknowledge the support of the UK Engineering and Physical Sciences Research Council under projects EP/J010065/1, EP/L000407/1 and EP/M011054/1.

\bibliographystyle{plain}
\bibliography{tidal-turbines}

\begin{thebibliography}{10}

\bibitem{Barnett_etal_Submitted}
G.~L. Barnett, S.~W. Funke, and M.~D. Piggott.
\newblock {Hybrid global-local optimisation algorithms for the layout design of
  tidal turbine arrays}.
\newblock {\em {Renewable Energy}}, Submitted.

\bibitem{Byrd_etal_1995}
R.~H. Byrd, P.~Lu, J.~Nocedal, and C.~Zhu.
\newblock {A Limited Memory Algorithm for Bound Constrained Optimization}.
\newblock {\em {SIAM Journal on Scientific Computing}}, 16(5):1190--1208, 1995.

\bibitem{Culley_etal_2014}
D.~M. Culley, S.~W. Funke, S.~C. Kramer, and M.~D. Piggott.
\newblock {A hierarchy of approaches for the optimal design of tidal turbine
  farms}.
\newblock In {\em {Proceeedings of the 5th International Conference on Ocean
  Energy}}, 2014.

\bibitem{Culley_etal_2015}
D.~M. Culley, S.~W. Funke, S.~C. Kramer, and M.~D. Piggott.
\newblock {Tidal stream resource assessment through optimisation of array
  design with quantification of uncertainty}.
\newblock In {\em {Proceeedings of the European Wave and Tidal Energy
  Conference (EWTEC) 2015}}, 2015.

\bibitem{Culley_etal_2016}
D.~M. Culley, S.~W. Funke, S.~C. Kramer, and M.~D. Piggott.
\newblock {Integration of cost modelling within the micro-siting design
  optimisation of tidal turbine arrays}.
\newblock {\em {Renewable Energy}}, 85:215--227, 2016.

\bibitem{Deardorff_1970}
J.~Deardorff.
\newblock {A numerical study of three-dimensional turbulent channel flow at
  large Reynolds numbers}.
\newblock {\em Journal of Fluid Mechanics}, 41(2):453--480, 1970.

\bibitem{Deardorff_1971}
J.~W. Deardorff.
\newblock {On the magnitude of the subgrid scale eddy coefficient}.
\newblock {\em {Journal of Computational Physics}}, 7(1):120--133, 1971.

\bibitem{Divett_etal_2013}
T.~Divett, R.~Vennell, and C.~Stevens.
\newblock {Optimization of multiple turbine arrays in a channel with tidally
  reversing flow by numerical modelling with adaptive mesh}.
\newblock {\em {Philosophical Transactions of the Royal Society A}},
  371(1985):1471--2962, 2013.

\bibitem{Divett_etal_2014}
T.~Divett, R.~Vennell, and C.~Stevens.
\newblock {Channel scale optimisation of large tidal turbine arrays in packed
  rows using large eddy simulations with adaptive mesh}.
\newblock In {\em {Proceeedings of the 2nd Asian wave tidal energy conference,
  Tokyo}}, 2014.

\bibitem{Divett_etal_2016}
T.~Divett, R.~Vennell, and C.~Stevens.
\newblock {Channel-scale optimisation and tuning of large tidal turbine arrays
  using LES with adaptive mesh}.
\newblock {\em {Renewable Energy}}, 86:1394--1405, 2016.

\bibitem{Farrell_etal_2013}
P.~E. Farrell, D.~A. Ham, S.~W. Funke, and M.~E. Rognes.
\newblock {Automated derivation of the adjoint of high-level transient finite
  element programs}.
\newblock {\em {SIAM Journal on Scientific Computing}}, 35(4):C369--C393, 2013.

\bibitem{Flather_1976}
R.~A. Flather.
\newblock {A tidal model of the northwest European continental shelf}.
\newblock {\em {Memoires de la Soci\'{e}t\'{e} Royale des Sciences de
  Li\`{e}ge}}, 10(6):141--164, 1976.

\bibitem{Funke_2012}
S.~W. Funke.
\newblock {\em {The automation of PDE-constrained optimisation and its
  applications}}.
\newblock PhD thesis, Imperial College London, 2012.

\bibitem{FunkeFarrell_Submitted}
S.~W. Funke and P.~E. Farrell.
\newblock {A framework for automated PDE-constrained optimisation}.
\newblock {\em {ACM Transactions on Mathematical Software}}, Submitted.

\bibitem{Funke_etal_2014}
S.~W. Funke, P.~E. Farrell, and M.~D. Piggott.
\newblock {Tidal turbine array optimisation using the adjoint approach}.
\newblock {\em {Renewable Energy}}, 63:658--673, 2014.

\bibitem{Funke_etal_Submitted}
S.~W. Funke, S.~C. Kramer, and M.~D. Piggott.
\newblock {Design optimisation and resource assessment for tidal-stream
  renewable energy farms using a new continuous turbine approach}.
\newblock Submitted.

\bibitem{GeuzaineRemacle_2009}
C.~Geuzaine and J.-F. Remacle.
\newblock {Gmsh: A 3-D finite element mesh generator with built-in pre- and
  post-processing facilities}.
\newblock {\em International Journal for Numerical Methods in Engineering},
  79(11):1309--1331, 2009.

\bibitem{LarsonBengzon_2013}
M.~G. Larson and F.~Bengzon.
\newblock {\em {The Finite Element Method: Theory, Implementation, and
  Applications}}.
\newblock Springer, 2013.

\bibitem{Liek_2000}
G.~A. Liek.
\newblock {Horizontal Large Eddy Simulation using Delft2D-MOR: the influence of
  large horizontal eddies on the shape, depth and extent of scour holes}.
\newblock Master's thesis, {TU Delft}, 2000.

\bibitem{Logg_etal_2012}
A.~Logg, K.-A. Mardal, G.~N. Wells, et~al.
\newblock {\em {Automated Solution of Differential Equations by the Finite
  Element Method}}.
\newblock Springer, 2012.

\bibitem{Nash_etal_2015}
S.~Nash, A.~I. Olbert, and M.~Hartnett.
\newblock {Towards a Low-Cost Modelling System for Optimising the Layout of
  Tidal Turbine Arrays}.
\newblock {\em {Energies}}, 8:13521--13539, 2015.

\bibitem{Smagorinsky_1963}
J.~Smagorinsky.
\newblock {General Circulation Experiments with the Primitive Equations}.
\newblock {\em Monthly Weather Review}, 91(3):99--164, 1963.

\bibitem{Zhu_etal_1997}
C.~Zhu, R.~H. Byrd, P.~Lu, and J.~Nocedal.
\newblock {Algorithm 778: L-BFGS-B: Fortran subroutines for large-scale
  bound-constrained optimization}.
\newblock {\em {ACM Transactions on Mathematical Software (TOMS)}},
  23(4):550--560, 1997.

\end{thebibliography}

\end{document}